\documentclass[aps,prl,twocolumn,amsmath,amssymb,superscriptaddress,citeautoscript,longbibliography,floatfix]{revtex4-1}

\usepackage[allow-number-unit-breaks]{siunitx}
\usepackage{graphicx}% Include figure files
\usepackage{dcolumn}% Align table columns on decimal point
\usepackage{bm}
\usepackage{color}
\usepackage{mathrsfs}
\usepackage{amsfonts}
\usepackage{multirow}
\usepackage{natbib}

\usepackage{epsf}
\usepackage{epsfig}
\usepackage{epstopdf}
\usepackage{amsmath}

\begin{document}

\title{Coherence Control of Directional Nonlinear Photocurrent \\ in Spatially Symmetric Systems}

\author{Yongliang Shi}
\affiliation{Center for Spintronics and Quantum System, State Key Laboratory for Mechanical Behavior of Materials, Xi’an Jiaotong University, Xi’an, 710049, China}

\author{Jian Zhou}\email{Email: jianzhou@xjtu.edu.cn}
\affiliation{Center for Alloy Innovation and Design, State Key Laboratory for Mechanical Behavior of Materials, Xi’an Jiaotong University, Xi’an, 710049, China}

\begin{abstract}
The interplay between crystal symmetry and its optical responses is at the heart of tremendous recent advances in light-matter interactions and applications. Nonlinear optical processes that produce electric currents, for example bulk photovoltaic (BPV) effect, require inversion symmetry broken materials, such as ferroelectrics. In the current work, we demonstrate that such BPV current could be generated in centrosymmetric materials with excitation of out-of-equilibrium coherent phonons. This is much different from the generally studied static or thermally excited states. We show that depending on the oscillating phase factor of the coherent phonon, uni-directional static electric current can be observed, in addition to some terahertz alternating currents. We also generalize the conventional injection charge current into angular momentum (spin and orbital) degrees of freedom, and demonstrate spin and orbital BPV photocurrents under coherent phonons. Our findings open the pathway to exploring the exotic phonon-photon-electron coherent interactions in quantum materials.

\end{abstract}

\maketitle
\subsection{Introduction}
Light-matter interaction is one of the key routes to manipulating electrons and converting light energy into electrical energy. Conventional light energy harvesting and conversion into electric current relies on semiconductor heterostructures (such as p-n junctions), in which electrons and holes can be separated spatially under a built-in electric field. This requirement can be eliminated when one uses a single crystal with non-centrosymmetric structure ($\mathcal{P}$-broken), and light illumination could generate direct electric current, known as bulk photovoltaic (BPV) effect~\cite{Fridkin01CR,Aversa95PRB}. The inversion asymmetry yields anharmonic electric field potential, leading to a collective motion of electrons under light irradiation (when photon energy $\hbar\omega$ is typically larger than semiconductor bandgap $E_g$). Here, $\mathcal{P}$-broken is an acknowledged condition to generate bias-free photocurrent.
\par   
The requirement of $\mathcal{P}$-broken limits the candidate material set to produce photocurrents (including shift current and injection current, depending on the crystal symmetry and light polarization). Both shift current and injection current processes have been widely observed in various materials, including ferroelectrics, quantum wells, organic crystals, and two-dimensional (2D) interfaces. One has to note that most previous works focus on ferroelectrics in their static or thermally equilibrium state, which are inherently $\mathcal{P}$-broken~\cite{Gong18PRL,Tan16npjCM}. In this work, we suggest that injection photocurrent can be triggered in spatially symmetric materials with coherent phonon oscillation excited using a THz pulse irradiation. We perform first-principles density functional theory calculations~\cite{Hohenberg64PR,Kohn65PR,Kresse96PRB,Perdew96PRL,Blochl94PRB} and discuss a spatially centrosymmetric system, 2D phosphorene ($\alpha$-P) monolayer~\cite{Liu14AN} to illustrate our theory. The 2D materials are advantageous for their easy optical access and manipulation~\cite{Zhou18NL,Guo21JPCL,Xu21PRB}, owing to their ultrathin nature. In addition, the direct optical absorption is much reduced in 2D thin films, hence the hot carrier mobility drift current can be marginal, yielding better experimental observations. We show that the zone-center ($q=0$) infrared (IR) active phonon can be excited under an THz laser pulse with intermediate fluence, which brings coherent odd-parity mode oscillations. Under such dynamic state, circularly polarized light (CPL) irradiation could produce a nonzero unidirectional propagating current, according to injection photocurrent mechanism. One has to note that even though the average displacements of such coherent phonon vibration averages to be zero, which corresponds to the static centrosymmetric geometry, the photon irradiation induced injection current does not vanish and contains a static current component. Such counter-intuitive phenomenon arises due to the injection current microscopic mechanism, which reflects the velocity mismatch between the valence and conduction bands. The direction of such photocurrent can be determined by the phonon vibration phase when the CPL irradiates, thus a coherence control of charge current can be realized~\cite{Zhao06PRL}. Furthermore, we propose that under linearly polarized light (LPL) irradiation, one could generate angular momentum (AM) related photocurrent (both spin BPV current and orbital BPV current), which has similar injection current nature. This generalizes the photo-induced nanoelectronics into spin and orbital AM degrees of freedom, which carries magnetic moment information. Therefore, dynamic but nonzero photocurrent with tunable transport direction (rather than a pure THz alternating current which averages to be zero over time) and with charge, spin, and orbital information can be produced via a synergistic cooperation of coherent phonon and second order nonlinear injection BPV approaches. Such mechanism can be extended to other spatially (vertical mirror) symmetric materials, such as H-MoS$_2$ monolayer~\cite{inversionsym19PRL}.

\begin{figure}[t]
    \centering
    \includegraphics[width=0.45\textwidth]{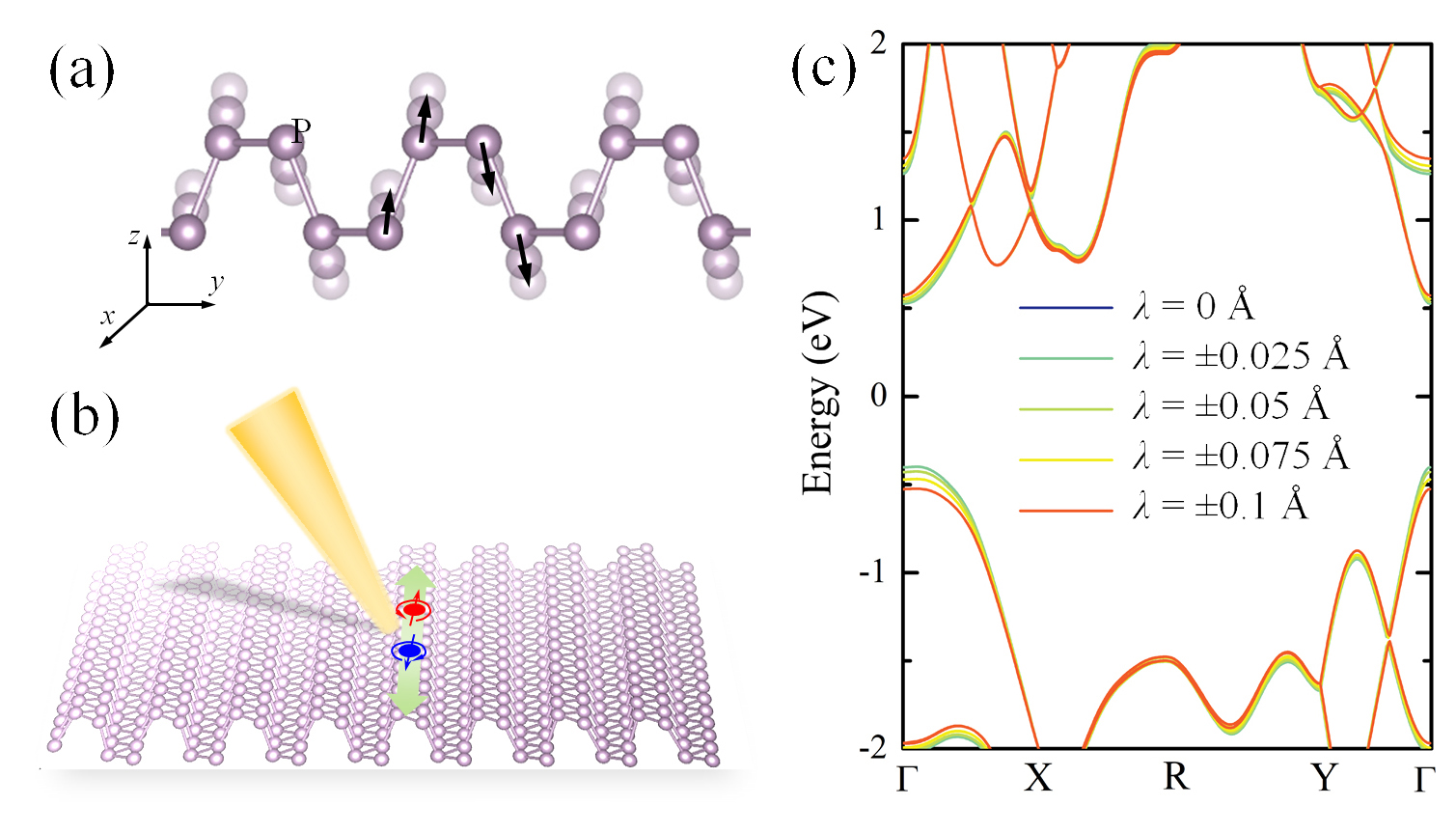}
    \caption{(a) THz optics irradiation inducing coherent IR-active phonon vibration in $\alpha$-P monolayer. The mode displacement amplitude is enlarged for clarity reason. (b) Nonlinear injection photocurrent move along the $x$-(zigzag) direction. (c) Quasi-static band dispersion variation under $\lambda$.}
    \label{fig:p}
\end{figure}

\par
According to nonlinear optics theory, once a CPL irradiates onto a nonmagnetic ($\mathcal{T}$-symmetric) material, its helicity phase generates injection current that is proportional to the interband Berry curvature (usually known as circular photogalvanic effect~\cite{Ji19NM}). In this process, light excites carriers into velocity-asymmetrical states in the Brillouin zone (BZ) and produces a net charge current~\cite{Sipe20PRB,Fregoso19PRB,Wang19SciAdv,Holder20PRR,Taghizadeh17PRB}. The photocurrent conductivity $\sigma(0;\omega,-\omega)$ grows linearly with time, and its rate is
\begin{equation}\label{eq:IC}
\begin{split}
\eta_{bc}^a(\omega)&=\frac{d}{dt}\sigma_{bc}^a(0;\omega,-\omega)
=-\frac{\pi e^3}{2\hbar^2}\int_{BZ}\frac{d^2\bm{k}}{(2\pi)^2} \\
&\times\mathrm{Im}\sum_{n,m}f_{nm}\Delta_{nm}^a[r_{mn}^b,r_{nm}^c]\delta(\omega_{mn}-\omega)
\end{split}
\end{equation}
Here $a,b,c$ indicate 2D Cartesian indices $x$ and $y$. $f_{nm}=f_n-f_m$ and $\Delta_{nm}^a=v_{nn}^a-v_{mm}^a$ are the occupation and velocity difference between band-$n$ and $m$, respectively. Interband transition dipole is $r_{mn}^b=\langle m|r^b|n\rangle$. Note that all explicit $\bm{k}$-dependence of quantities are omitted. The commutator $[r_{mn}^b,r_{nm}^c]=r_{mn}^br_{nm}^c-r_{mn}^cr_{nm}^b=-i\epsilon_{abc}\Omega_{mn}^a$ corresponds to interband Berry curvature. We use Lorentzian function to represent the Dirac $\delta$-function, with a broadening factor of $0.02\,\mathrm{eV}$. It could phenomenologically incorporate the temperature, disorder, and scattering effects. The total current evolves as $j_{\mathrm{inj}}^a=2\sigma_{bc}^a(0;\omega,-\omega)E_b(\omega)E_c(-\omega)$, where $E_i$ indicates optical electric field component. According to Eq.~\ref{eq:IC}, the different velocities at the valence and conduction bands produces the net injection photocurrent. For the 2D BZ integration which lacks the out-of-plane $k_z$ component, the calculated $\eta$ (and $\sigma$) includes a surplus length unit (nm or \AA) compared with that in conventional 3D crystals. In order to eliminate such difference and adopt the 3D unit systems, we rescale them by utilizing an effective thickness of the system, $\eta^{\mathrm{eff}}=\eta^{\mathrm{2D}}/d$. Here $d$ refers to the monolayer effective thickness (taken to be $6\,\mathrm{\AA}$ in the current work), and the superscripts ``2D" and ``eff" indicate 2D values and the rescaled effective results. This is based on a parallel capacitor model and is widely used previously~\cite{Zhou18NL,Laturia18npj2dm,Jiang17PRL,Mu21npjCM}. In this work, we report these rescaled values.
\subsection{Results}
\par
\textit{Coherent vibration of $\alpha$-P monolayer and injection current rate under CPL.} According to symmetry arguments, we limit our discussion on the zone-centered phonon modes (unit cell wavelength coherent oscillation), which can be effectively excited via THz optics (Supplemental Material, SM). In $\alpha$-P monolayer, the only one is $B_{1u}$ (at $4.2\,\mathrm{THz}=140\,\mathrm{cm}^{-1}$), which breaks $\mathcal{P}$ and introduces dynamic in-plane polarization ($P_y(t)\neq 0$, see SM). This mode mainly displaces P atoms out-of-plane, as shown in Fig.~\ref{fig:p}(a). We find that the anharmonicity of $B_{1u}$ is very weak (see SM), suggesting its small damping effect and long lifetime during operation. Now we displace atoms according to their phonon eigenvectors, and use $\lambda$ to measure the total displacements of all four atoms in a unit cell. When $\lambda=\pm0.1\,\mathrm{\AA}$, the energy gain is $10\,\mathrm{meV}$ per atom. This corresponds to a THz laser pulse with pump fluence of $0.25\,\mathrm{mJ/cm^2}$ centered at $4.2\,\mathrm{THz}$ (see SM). The positive and negative $\lambda$ correspond to spatially opposite images, i.e., atomic coordinates $\mathbf{x}_i^{\lambda}=-\mathbf{x}_i^{-\lambda}$ for each atom $i$. Hence, they show same electronic energy band structure [Fig.~\ref{fig:p}(c)].

\begin{figure*}[bt]
    \centering
    \includegraphics[width=0.75\textwidth]{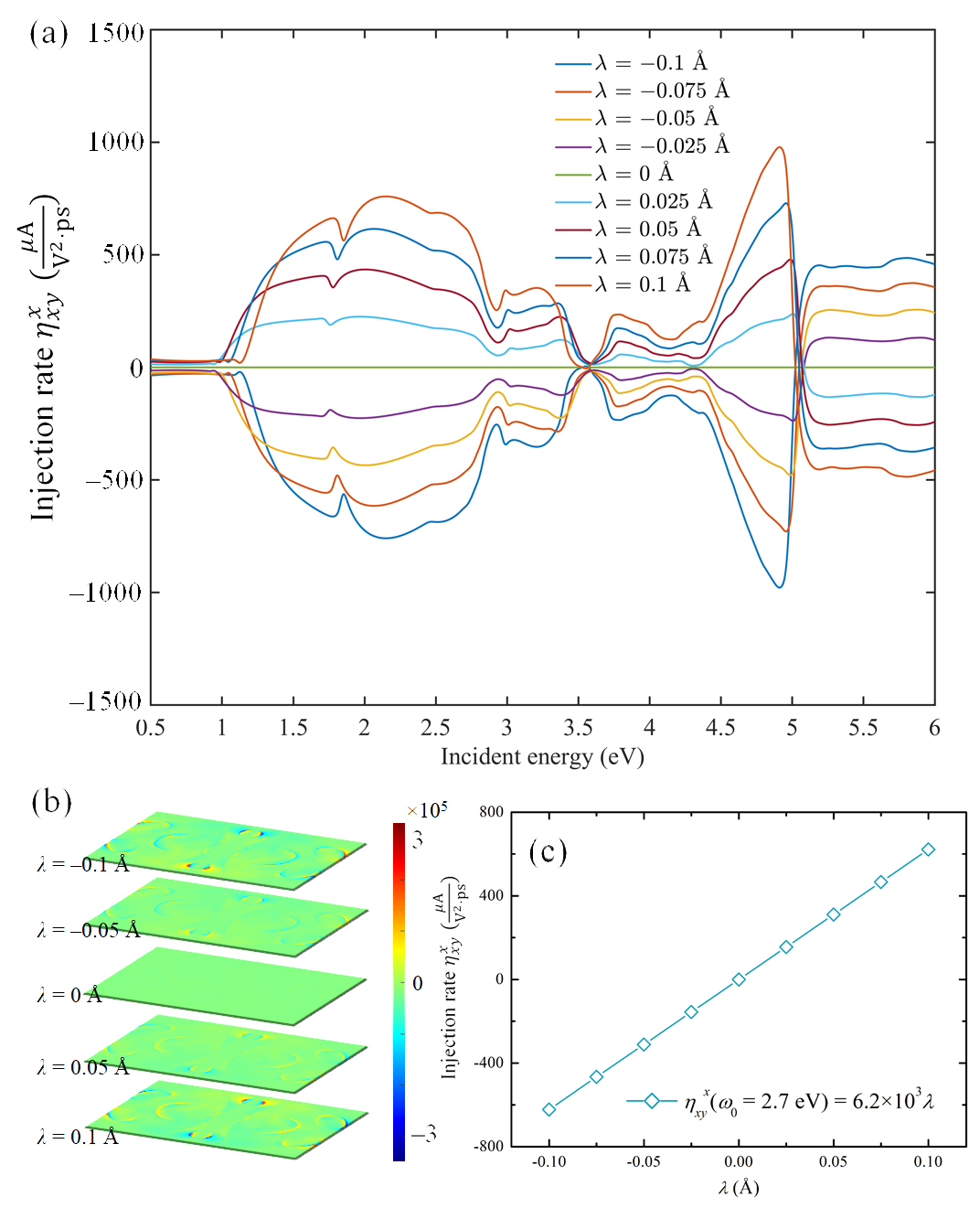}
    \caption{(a) Injection photocurrent rate $\eta_{xy}^x$ of $\alpha$-P in various $\lambda$ states. (b) $k$-resolved contribution of injection photocurrent rate at $\omega_0=2.7\,\mathrm{eV}$. (c) Linear fitting of $\eta_{xy}^x(\omega_0)$ with $\lambda$.}
    \label{fig:cpl}
\end{figure*}

\begin{figure*}[htb]
    \centering
    \includegraphics[width=0.85\textwidth]{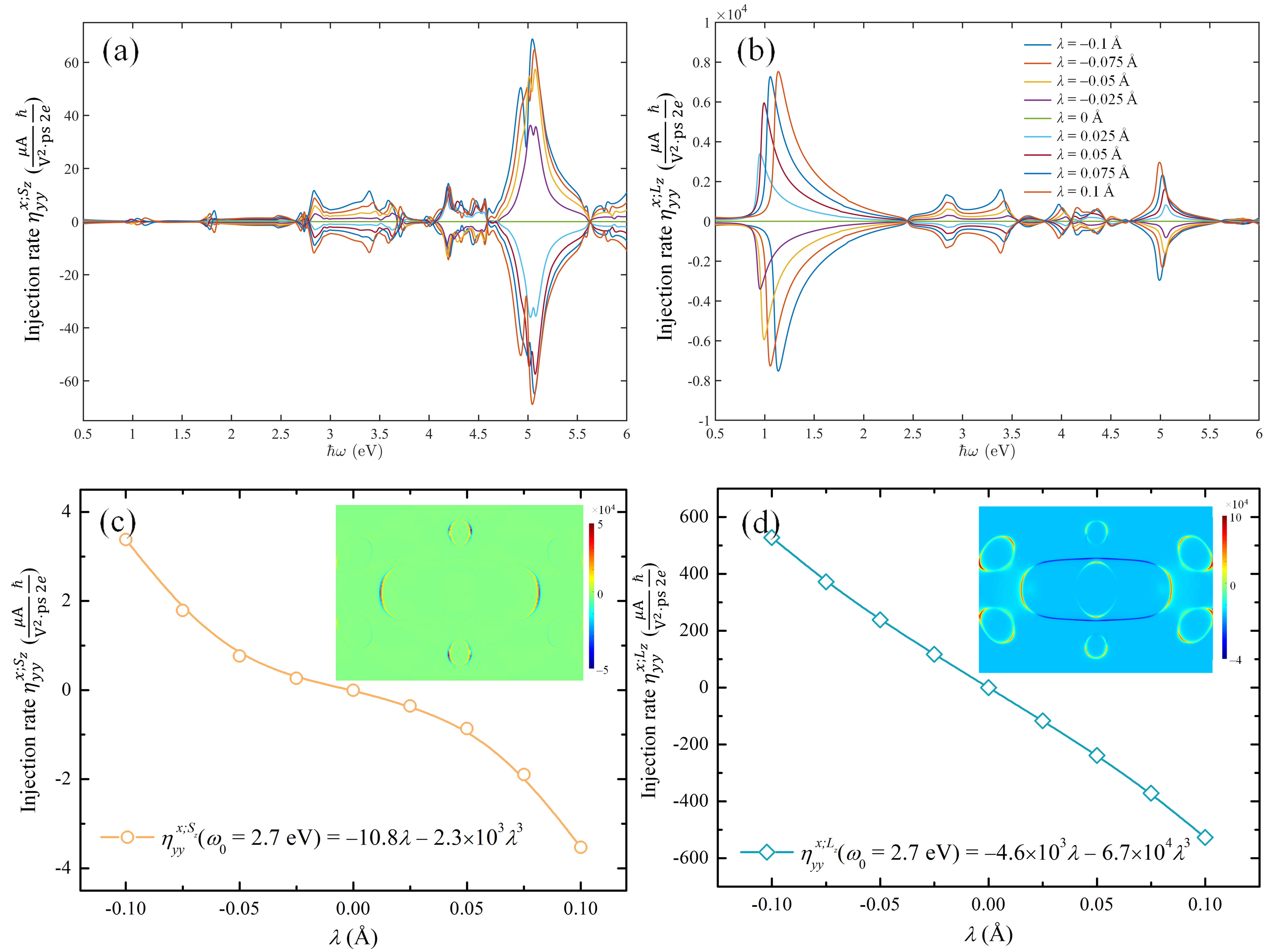}
    \caption{(a) Spin and (b) orbital photocurrent rate under different $\lambda$. (c) and (d) show their $\lambda$-dependence at $\omega_0=2.7\,\mathrm{eV}$. Insets show the $k$-resolved contributions at $\lambda=-0.1\,\mathrm{\AA}$.}
    \label{fig:amlpl}
\end{figure*}

\par
The coherent $B_{1u}$ reduces the space group from static $Pmna$ to $Pmn2_1$, which breaks the $\mathcal{M}_y$ mirror symmetry but preserves $\mathcal{M}_x$. According to our symmetry analysis (see SM), dynamic injection current then emerge under CPL and would flow along the $x$-direction. Here we will only discuss $\eta_{xy}^x$ since $\eta_{yx}^x=-\eta_{xy}^x$. Since the ionic vibration periodic is on a few hundreds of picosecond, one can safely assume that electrons effectively experience a quasi-static potential field under different $\lambda$. Figure~\ref{fig:cpl}(a) shows the injection current rate as a function of different displacement. One clearly sees that positive and negative $\lambda$ give opposite photoconductivity rates ($\eta_{xy}^x(\lambda)=-\eta_{xy}^x(-\lambda)$), even though their band dispersions are the same. We also plot the $k$-resolved integrand of Eq.~\ref{eq:IC} at a selected optical incident energy $\omega_0=2.7\,\mathrm{eV}$, which shows large contribution near the $\Gamma-Y$ paths [Fig.~\ref{fig:cpl}(b)]. This distribution is a bit different from the joint density of states (see SM), suggesting anisotropic interband transition contribution $\langle m|\bm{r}|n\rangle$ at different $\bm{k}$. We then fit $\eta_{xy}^x(\omega_0)$ with respect to $\lambda$ by linear regression $\eta_{xy}^x(\omega_0)=C_1\lambda+o(\lambda^2)$ [Fig.~\ref{fig:cpl}(c)],  which gives $C_1=6.2\times 10^3\,\frac{\mu\mathrm{A}}{\mathrm{V}^2\mathrm{\AA}\cdot\mathrm{ps}}$.

\par
\textit{Angular momentum current under LPL irradiation.} Photocurrent generated by CPL also includes LPL shift current feature, while LPL irradiation does not have such problem~\cite{Ahn20PRX}. Hence, next we consider LPL induced injection current. For time-reversal symmetry systems, we have recently demonstrated that LPL could generate AM photocurrents, showing injection current feature~\cite{Mu21npjCM,Xu21NC}. Photo-induced AM currents include both spin and orbital currents. They indicate that electrons carrying opposite spin and orbital moments are flowing oppositely. Note that orbital angular momentum is a largely overlooked degree of freedom that has attracted increasingly attention recently~\cite{Bernevig05PRL}. We evaluate the spin (and orbital) photocurrent conductivity rate~\cite{Bhat05PRL,Mu21npjCM,Fei20Arxiv,Fei21NL}
\begin{equation}\label{eq:AOC}
\begin{split}
\eta_{bc}^{a;\hat{O}_z}(\omega)&=\frac{d}{dt}\sigma_{bc}^{a;\hat{O}_z}(0;\omega,-\omega)=-\frac{\pi e^3}{2\hbar^2}\int_{BZ}\frac{d^2\bm{k}}{(2\pi)^2} \\
&\times\mathrm{Re}\sum_{n,m}f_{nm}j_{nm}^{a;\hat{O}_z}\{r_{mn}^b,r_{nm}^c\}\delta(\omega_{mn}-\omega)
\end{split}
\end{equation}
Here $j_{nm}^{a;O_z}$ denotes the difference of angular momentum current for band-$n$ and $m$, and $j^{a;\hat{O}_z}=\{v_a,\hat{O}_z\}/2e,\,(\hat{O}=\hat{S},\hat{L})$~\cite{Shi06PRL,Cysne21PRL}. The $\{r_{mn}^b,r_{nm}^c\}$ refers to anticommutation between $r_{mn}^b$ and $r_{nm}^c$. Here we focus on the spin-$z$ and orbital-$z$ components, since they are uni-directionally pointing along the out-of-plane directions in most $k$-points (singly degenerate)~\cite{Pan19PRAppl,Mu21npjCM}. This is because $\bm{S}$ and $\bm{L}$ are pseudovectors and the $Pmn2_1$ contains glide mirror operation $\{\mathcal{M}_z\vert\frac{1}{2}\frac{1}{2}0\}$.

\par
Figure~\ref{fig:amlpl} shows our calculated nonlinear spin and orbital photocurrent rates under $y$-LPL for different $\lambda$. The $x$-LPL responses are plotted in SM. Similar as in the $\eta_{xy}^x$ case, the static $\mathcal{P}$-symmetric system constraints that $\eta_{yy}^{x;S_z}(\lambda=0)=0$ and $\eta_{yy}^{x;L_z}(\lambda=0)=0$. Positive and negative $\lambda$ gives opposite values for the AM photocurrent rates. One has to note that under LPL, there is no shift current along $x$, due to presence of $\mathcal{M}_x$. Thus, there is no charge current mixture in the current case (pure spin and orbital currents). We observe that the orbital photocurrent has much larger values than the spin photocurrent, owing to small spin-orbit coupling in P. This is in contrast with the common picture that orbital moments are usually quenched in magnetic systems under strong and isotropic crystal field effect, where magnetisms are mainly from spin degree of freedom. We also fit the these photocurrent rates at $\omega_0=2.7\,\mathrm{eV}$ by odd power polynomial function using $\lambda$, $\eta_{yy}^{x;S_z}(\omega_0)=C_1^S\lambda+C_3^S\lambda^3+o(\lambda^4)$ and $\eta_{yy}^{x;L_z}(\omega_0)=C_1^L\lambda+C_3^L\lambda^3+o(\lambda^5)$. Our results are $C_1^S=-10.8$, $C_3^S=-2.3\times 10^3$, $C_1^L=-4.6\times 10^3$, and $C_3^L=-6.7\times 10^4$. The units of $C_1^{S,L}$ and $C_3^{S,L}$ are $\frac{\mu\mathrm{A}}{\mathrm{V}^2\mathrm{\AA}\cdot\mathrm{ps}}\frac{\hbar}{2e}$ and $\frac{\mu\mathrm{A}}{\mathrm{V}^2\mathrm{\AA}^3\cdot\mathrm{ps}}\frac{\hbar}{2e}$, respectively. Both of them show strong nonlinear $\lambda^3$ relationship due to the interplay of bandgap and wavefunction variations under $B_{1u}$.

\par
\textit{Time-dependent injection charge, spin, and orbital currents.} We now integrate $\eta$ in time to explore these second order photoconductivities under coherent $B_{1u}$ oscillation
\begin{equation}\label{eq:sigma}
\frac{d\sigma_{\mathrm{inj}}(t)}{dt}=\eta(\omega_0,\lambda(t))
\end{equation}
where $\lambda(t)=\lambda_0\sin(2\pi\omega_{\mathrm{IR}}t+\phi)$ is the $B_{1u}$ mode vibration equation of motion with amplitude of $\lambda_0=0.1\,\mathrm{\AA}$ and frequency $\omega_\mathrm{IR}=4.2\,\mathrm{THz}$. $\phi\in(-\pi,\pi]$ is the vibration phase at the initial state ($t=0$). The initial condition is $\sigma_{\mathrm{inj}}(t=0)=0$.

\par
\begin{figure*}[t]
    \centering
    \includegraphics[width=0.75\textwidth]{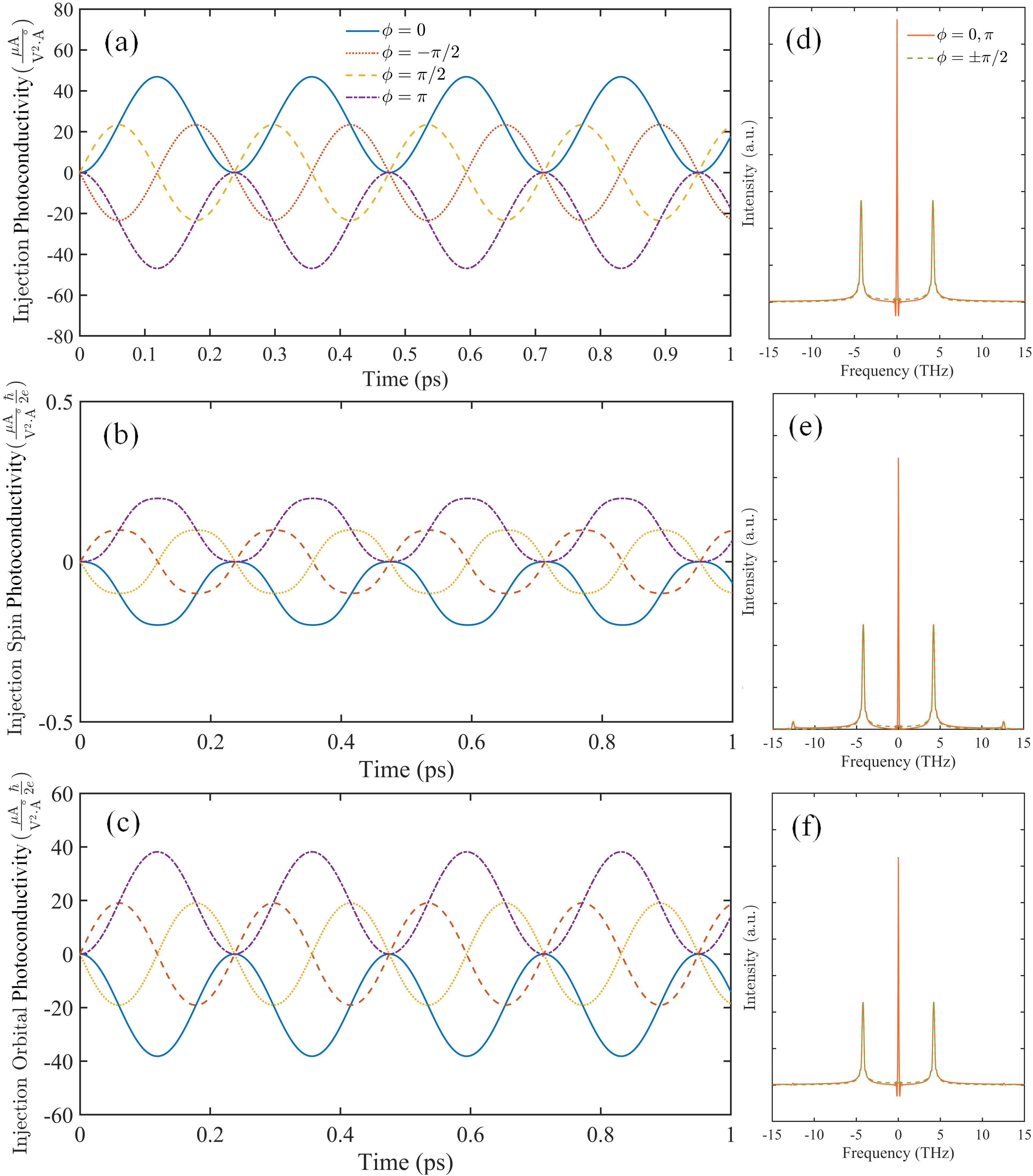}
    \caption{Time evolution of (a) injection photoconductivity under CPL, (b) spin and (c) orbital photoconductivity under LPL. The current propogation strongly depends on phase $\phi$. In (d)$-$(f) we show frequency distributions of (a)$-$(c) photocurrents by Fouier transformations. Incident photon energy of $\omega_0=2.7\,\mathrm{eV}$ is used.}
    \label{fig:timeint}
\end{figure*}

As shown in Fig.~\ref{fig:timeint}(a)$-$(c), the photocurrent oscillates around an average value, which is determined by $\phi$. In the coherent vibration case, the $\phi$ is the same in different unit cells. When $\phi=\pm\pi/2$, one always obtains an oscillating (charge, spin, and orbital) current. The time average of such current is zero $-$ indicating a pure alternating current (with propagating directions opposite for $\phi=\pm\pi/2$). When we perform Fourier transformation of such currents [dashed curves in Fig.~\ref{fig:timeint}(d)$-$(f)], it shows sharp peak at $\omega_{\mathrm{IR}}=4.2\,\mathrm{THz}$. For the spin and orbital photocurrents, addtional peaks at $3\omega_{\mathrm{IR}}=12.6\,\mathrm{THz}$ can be seen, consistent with the $\lambda^3$-dependence of $\eta_{yy}^{x;O_z}(\omega_0)$ in these cases. Depending on the incident photon energy, such non-linearity may be tuned. Remarkably, when $\phi=0$ and $\phi=\pi$, one can see that in additional to the oscillatory nature of these currents ($\omega_{\mathrm{IR}}$ and $3\omega_{\mathrm{IR}}$ for AM photocurrent peaks), they also contain large static current (or AM currents) features with $0\,\mathrm{THz}$ contributions. This indicates that one could observe a \textit{uni-directional} injection photocurrent, the magnitude of which wiggles but it always flows toward one direction. Depending on the sign of $C$ (and $C^S$, $C^L$), the photocurrent can be positive or negative. The $\phi=0$ and $\phi=\pi$ give opposite propagating currents. Note that the time integral cannot be taken for arbitrarily long time, as electron relaxation time $\tau$ is always finite\cite{Xu21NC,Barik20PRB} (but usually longer than a few picoseconds).

\par
\begin{figure}[b]
    \centering
    \includegraphics[width=0.45\textwidth]{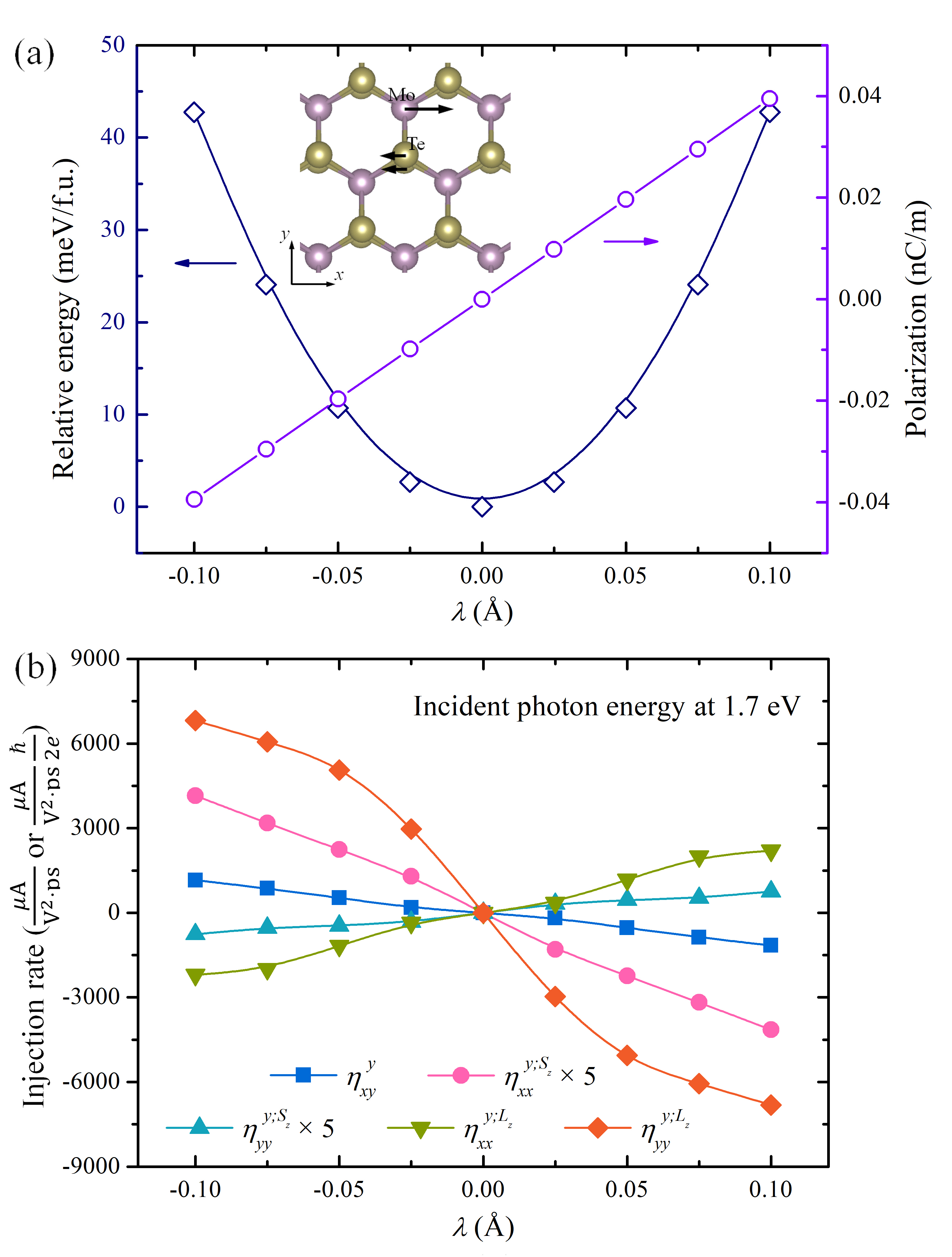}
    \caption{(a) IR-active mode (along $x$, zigzag direction) driven variation of total energy and electric polarization ($P_x$) in MoTe$_2$. Inset shows atomic geometry and vibration mode. (b) shows the variation of injection photocurrent rate under CPL (unit of $\frac{\mu\mathrm{A}}{\mathrm{V}^2\cdot\mathrm{ps}}$) and injection AM photocurrent rate under LPL (unit of $\frac{\mu\mathrm{A}}{\mathrm{V}^2\cdot\mathrm{ps}}\frac{\hbar}{2e}$). All currents propagate along $y$ (armchair).}
    \label{fig:mote2}
\end{figure}

\textit{Transition metal dichalcogenide monolayer case.} So far we have demonstrated that coherent IR-active oscillation could induce finite injection current in a static $\mathcal{P}$-symmetric system,  $\alpha$-P monolayer. In order to show that it is a universal effect, we now briefly discuss such oscillating injection current in another well-studied monolayer, transition metal dichalcogenide MoTe$_2$~\cite{Wang17PRB}. This system does not possess $\mathcal{P}$ at its static structure, but it includes a $\mathcal{M}_x$ vertical mirror plane [inset of Fig.~\ref{fig:mote2}(a)]. It symmetrically forbids CPL injection current and LPL AM photocurrent along $y$. In the unit-cell coherent vibrations, there is a doubly degenerate $E'$ mode (at $\omega_{\mathrm{IR}}'=7.1\,\mathrm{THz}$) being IR-active. It contains two vertical vibration eigenvectors, along $x$ and $y$ directions, respectively. They can be individually excited using linearly polarized THz pulses. Hence, we only focus on the $x$-mode which breaks $\mathcal{M}_x$ dynamically.

\par
We again consider vibrations with $\lambda$ between $-0.1\,\mathrm{\AA}$ and $0.1\,\mathrm{\AA}$. The total energy increases quadratically up to 43 meV/f.u. The system shows a finite electric polarization along zigzag direction, $P_x$, which linearly increases with $\lambda$. Then we consider the injection photocurrents. When the ionic system is coherently excited ($E_x'$), the $\mathcal{M}_x$ symmetry constraints lift. From Fig.~\ref{fig:mote2} we plot their values at a chosen incident photon energy ($1.7\,\mathrm{eV}$). Nonlinear $\lambda$-dependence can be clearly observed in this case, especially for orbital and spin photocurrents. If we integrate the photocurrent rate as in Eq.~\ref{eq:sigma}, we obtain similar uni-directional injection photocurrent ($0\,\mathrm{THz}$ static current component), and the nonlinear contributions ($3\omega_{\mathrm{IR}}'$ and $5\omega_{\mathrm{IR}}'$ contributions) can be obtained (see SM).
\subsection{Discussion}
\par
Before concluding, we would like to note that the coherent IR-active vibration at $\Gamma$ is different from other ionic vibrations or motions, such as thermal excitations, Raman active modes, or BZ boundary modes. Due to equi-partition theorem, the thermally excited phonons are random in phase and do not possess a definite wavelength. Thus, there is no collective and coherent motion of ions under thermal excitation. The phase factor $\phi$ does not have a long range correlation in different unit cells. This will result in qualitatively similar behaviors as the static configuration, and both alternative (at $\omega_{\mathrm{IR}}$ and static injection currents vanish. Collective $\Gamma$-centered Raman modes and BZ boundary modes usually do not break the spatial symmetry $\mathcal{P}$ or $\mathcal{M}$. Hence, even though they would affect the electronic structure (band dispersion, etc) and can change the otherwise existing current values, the spatial symmetry arguments do not break. Then the oscillating effect will be different and may not be as strong as that in our current case. We will study their effects in a separate work. We also want to note that recently Schlipf and Giustino proposed dynamic Rashba-Dresselhaus (dRD) effect under coherent IR-active phonon oscillations~\cite{Schlipf21dRD}, where they showed that the dRD can be observed at a given time $t$. But in their case the dRD would diminish if time average is performed. In our proposal, on the contrary, we can obtain static (zero frequency) photocurrent component, which does not require an ultrafast dynamics observation setup. Of course, the recently developed ultrafast dynamic approaches could provide high resolution observation of the injection photocurrent~\cite{Luo21NM}, and would be useful for experimental verification and applications of our prediction.
\par
In our current cases, the $\alpha$-P and MoTe$_2$ monolayers possess out-of-plane glide and mirror symmetries, respectively. They forbid in-plane angular momentum components, since $\bm{L}$ and $\bm{S}$ are both axial vectors. However, one has to note that $L_z$ does not commute with the Hamiltonian as it is determined by crystal field effect. Hence, the orbital angular momentum is generally not a good quantum number. This would yield additional orbital torque effect\cite{Go20PRResearch} in the continuum equation of motion. Actually, similar problem occurs in quantum spin Hall insulator when Rashba spin-orbit coupling is present. Various approaches have been suggested to avoid such problem, for example, one can project $L_z$ onto eigenstate space via $\mathbb{L}_z=\sum_n |\psi_{n\bm{k}}\rangle\langle\psi_{n\bm{k}}|L_z|\psi_{n\bm{k}}\rangle\langle\psi_{n\bm{k}}|$\cite{Cysne21PRL,Phong19PRL,Go18PRL}, which corresponds to the diagonal terms in density matrix. Then the $[\mathbb{L}_z,\hat{H}]=0$. Our test calculations show that such a scheme gives qualitatively the same results as discussed here with the symmetry arguments still hold. But a well-defined orbital operator still requires theoretical efforts.
\par
The definition of angular momentum current is another open problem. When the out-of-plane spin-$z$ component is conserved, one may define the spin current as $j_{\mathrm{sp}}=j_{\uparrow}-j_{\downarrow}$, where $\uparrow$ and $\downarrow$ indicate spin-up and spin-down channels, respectively. However, in a general case, there are several spin current expressions. A widely-adopted approach (also in the current study) is $j_{\mathrm{sp}}^{a;S_z}=\{v_a,S_z\}$, which has been used to evaluate spin Hall effect and spin Seebeck effect, etc. On the other hand, it has also been proposed that a proper definition takes the form of $j_{\mathrm{sp}}^{a;S_z}=\frac{d(r_aS_z)}{dt}$. We have compared both these approaches and showed that they yield similar results with small differences\cite{Xu21NC}. In the current study, our focus is on the prediction of static injection current components induced by coherent phonon vibration, while such differences would not affect the main results.
\par
In this study, we use the quadratic recursive response theory to evaluate the injection current rate, which essentially corresponds to $\mathrm{Tr}(j^{(0)}\rho^{(2)})$, where $j^{(0)}$ is electric field $E$-independent current operator and $\rho^{(2)}$ is the second order perturbative density matrix operator. They can be calculated according to Kubo theory. Note that there are other contributions, such as self energy and vertex corrections\cite{Avdoshkin20PRL}, which cannot be simply included in the current density functional theory calculations and is not computed here. In addition, the $\mathrm{Tr}(j^{(1)}\rho^{(1)})$ contribution, where $j^{(1)}$ originates from anomalous velocity and is linearly dependent on electric field $E$, would vanish in such topologically trivial and insulating systems. The band edge dephasing time is calculated to be a few femtoseconds, increases with temperature reduces (see SM). This is consistent with the previous work~\cite{Zereshiki18NS}, and is sufficiently long for nonlinear optical process (absorbing and releasing photons during the two-photon-one-electron process).
\par
As discussed previously, bulk photovoltaic effect includes shift current and injection current, which belong to different microscopic mechanisms\cite{Tan16npjCM}. Taking charge current as an example (similar arguments hold for AM current), the shift current arises from the mismatch of wavefunction center position in the valence and conduction bands. On the other hand, injection current originates from the mismatch of velocity operator in the valence and conduction bands. Hence, the injection current \textit{rate} is more fundamental. Most previous studies evaluate the injection current for static configurations via $\sigma_{\mathrm{inj}}=\eta\tau$ where $\tau$ is electron relaxation lifetime, determined by disorder, temperature, impurities, and electron-phonon scattering, etc\cite{Wang19SciAdv,Xu21NC,Fei20Arxiv}. In this study we show that coherent vibration gives a time dependent rate $\eta(t)$, hence the ultimate injection current can be dynamic, nonzero, and even contains static components. The shift current, on the contrary, does not have such property. In realistic experimental setups, above bandgap photon absorption produces hot carriers (electrons and holes), and their drifts contribute to photoconductivities as well (e.g., ballistic current). This is not the situation discussed in the current work. One also has to note that damping effect would gradually suppress the static current component (see SM).
\par

\subsection{Conclusion}
\par
In summary, we apply first-principles calculations to show that $\Gamma$-centered coherent IR-active phonon oscillation that breaks spatial symmetry could trigger nonlinear injection photocurrent in otherwise silent systems. This photocurrent could be pure alternating or uni-directionally propagating, depending on the phase of the coherent phonon. Our work shows that unlike the widely studied static or thermally equilibrium conditions, the synergistic interplay between the coherent phonon, electron, and photon could lead to intriguing and exotic nonlinear optical responses in ultrafast dynamics. The system setup we propose here may serve as a future THz electromagnetic wave detector for the next generation communication.

\begin{acknowledgments}
This work is supported by the National Natural Science Foundation of China (NSFC) under Grant Nos. 11974270, 21903063, and 11904353. J.Z. acknowledges valuable discussions with H. Xu (MIT) and Y. Gao (USTC), and the computational resources provided by HPCC platform of Xi’an Jiaotong University.
\end{acknowledgments}

\bibliography{Mainbib}% Produces the bibliography via BibTeX.
\end{document}